\documentclass[prd,10pt,aps,showpacs,amsmath,amssymb,twocolumn,nofootinbib]{revtex4-1}

\usepackage[colorlinks]{hyperref}
\hypersetup{linkcolor=blue,citecolor=blue}

\usepackage{graphicx,amsfonts,amssymb,amsbsy}
\usepackage{amsmath,amsthm,latexsym}
\usepackage{pstricks}
\usepackage{slashed}
\usepackage{ulem}

\def\ua1{U_A(1)}

\begin{document}

\title{Surface tension of quark droplets in compact stars and in the early universe}

\author{A. G. Grunfeld$^{1,2}$}
\email{grunfeld@tandar.cnea.gov.ar}

\author{G. Lugones$^{3}$}
\email{german.lugones@ufabc.edu.br}

\affiliation{$^1$ CONICET, Godoy Cruz 2290, (C1425FQB) Buenos Aires, Argentina.}
\affiliation{$^2$ Departamento de F\'\i sica, Comisi\'on Nacional de Energ\'{\i}a At\'omica, Av. Libertador 8250, (1429) Buenos Aires, Argentina.} 
 \affiliation{$^3$ Universidade Federal do ABC, Centro de Ci\^encias Naturais e Humanas, Avenida dos Estados 5001 - Bang\'u, CEP 09210-580, Santo Andr\'e, SP, Brazil.}

\begin{abstract}
We review the role of the surface tension of quark matter droplets in astrophysical conditions, focusing specifically on the thermodynamic conditions prevailing in cold neutron stars (NSs), in hot lepton rich proto NSs, and in early universe conditions. We analyze quark matter in chemical equilibrium under weak interactions, which is relevant for understanding the internal composition of hybrid stars, as well as ``just deconfined" quark matter out of chemical equilibrium, which is the relevant thermodynamic state for describing the nucleation process of quark matter in NSs. We explore the role of temperature, density, trapped neutrinos, droplet size and magnetic fields within the multiple reflection expansion formalism (MRE). Quark matter is described within the frame of different effective models: the MIT bag model and the $SU(3)_f$ Nambu-Jona-Lasinio model (NJL), including color superconductivity, neutrino trapping and magnetic fields. We also analyze the deconfinement transition at vanishing chemical potential and finite temperature including the Polyakov loop.  We explore some astrophysical consequences of our results.
\end{abstract}


\maketitle

\section{Introduction}
The cores of massive NSs have densities that may favor the nucleation of small droplets of quark matter, that under appropriate conditions may grow converting a large part of the NS into a deconfined state.
The study of the surface tension of deconfined quark matter has attracted much attention recently \cite{Buballa2013Pinto2012Wen2010Palhares2010} because a detailed knowledge of it may contribute to a better comprehension of the physics of NS interiors. In fact, surface tension plays a crucial role in quark matter nucleation during the formation of compact stellar objects, because it determines the nucleation rate and the associated critical size of the nucleated drops \cite{doCarmo2013,Lugones2011}. It is also determinant in the formation of mixed phases at the core of hybrid stars which may arise only if the surface tension is smaller than a critical value of the order of tens of MeV/fm$^2$ \cite{Voskresensky2003Tatsumi2003Maruyama2007Endo2011}. Also, surface tension affects decisively the properties of the most external layers of a strange star which may fragment into a charge-separated mixture, involving positively-charged strangelets immersed in a negatively charged sea of electrons, presumably forming a crystalline solid crust \cite{Jaikumar2006}. This would happen below a critical surface tension which is typically of the order of a few MeV/fm$^2$ \cite{Alford2006}.

However, in spite of its key role for NS physics, the surface tension is still poorly known for quark matter. Early calculations gave rather low values, below $5  \, \mathrm{MeV/fm^2}$ \cite{Berger1987}, but larger values within  $10 - 50 \, \mathrm{MeV/fm^2}$ where used in further works about quark matter droplets in NSs \cite{Heiselberg1993Iida1998}. More recently, values around $\approx 30 \, \mathrm{MeV/fm^2}$ have been adopted for studying the effect of quark matter nucleation on the evolution of proto NSs \cite{Bombaci2007Bombaci2009}. However, values around  $\sim 300 \, \mathrm{MeV/fm^2}$ were suggested on the basis of dimensional analysis of the minimal interface between a color-flavor locked phase and nuclear matter \cite{Alford2001}. 

In this work we present results we obtained for finite size effects within the multiple reflection expansion (MRE) framework (for details on the MRE formalism see \cite{Madsen-dropKiriyama1Kiriyama2} and references therein). To describe quark matter we employ the NJL and the MIT bag models considering different scenarios that we describe below.

\section{Surface tension and deconfinement at protoneutron stars}

\begin{figure*}[t]
\hskip -8mm
\includegraphics[angle=0,scale=0.25]{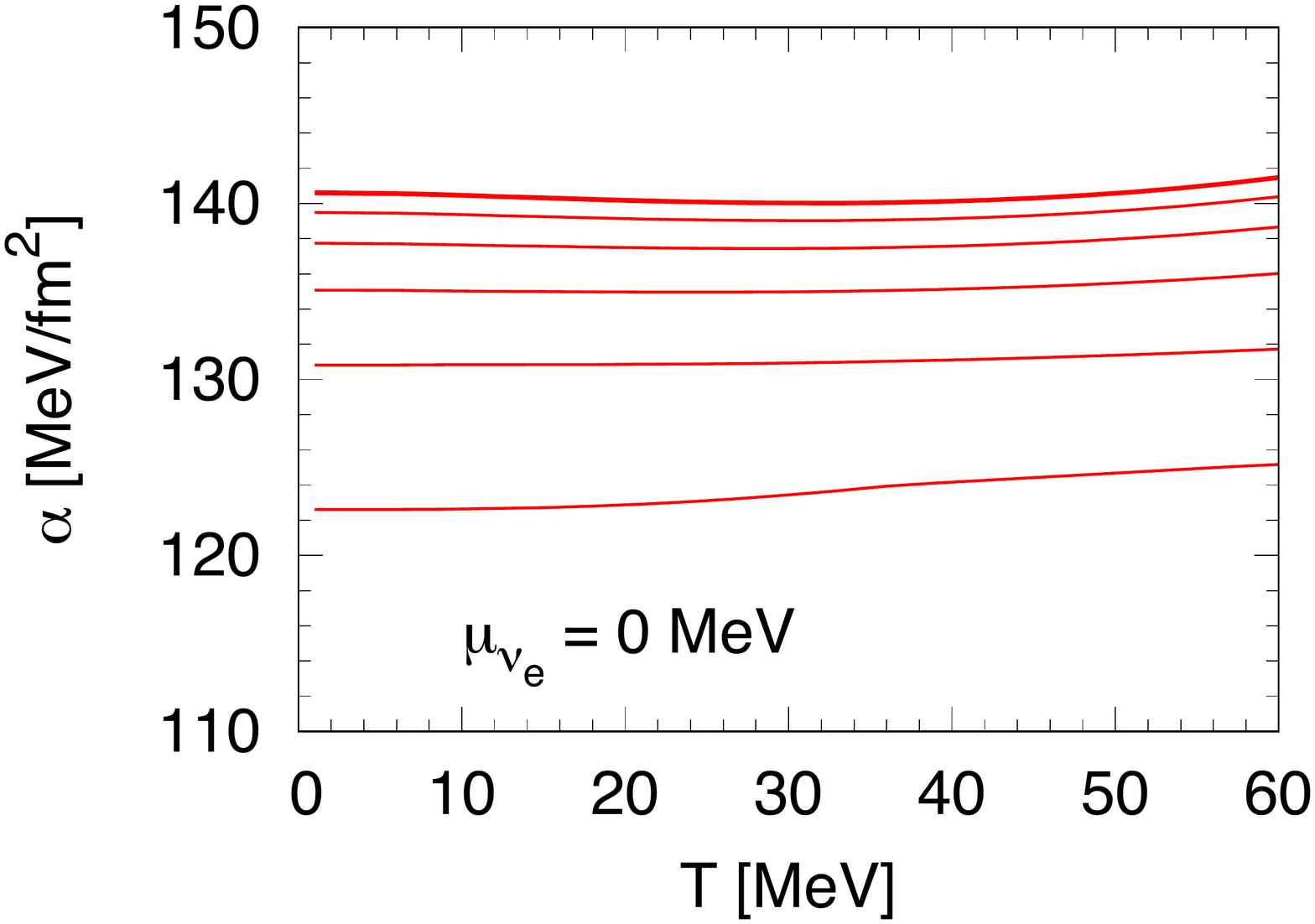}
\hskip -8mm
\includegraphics[angle=0,scale=0.25]{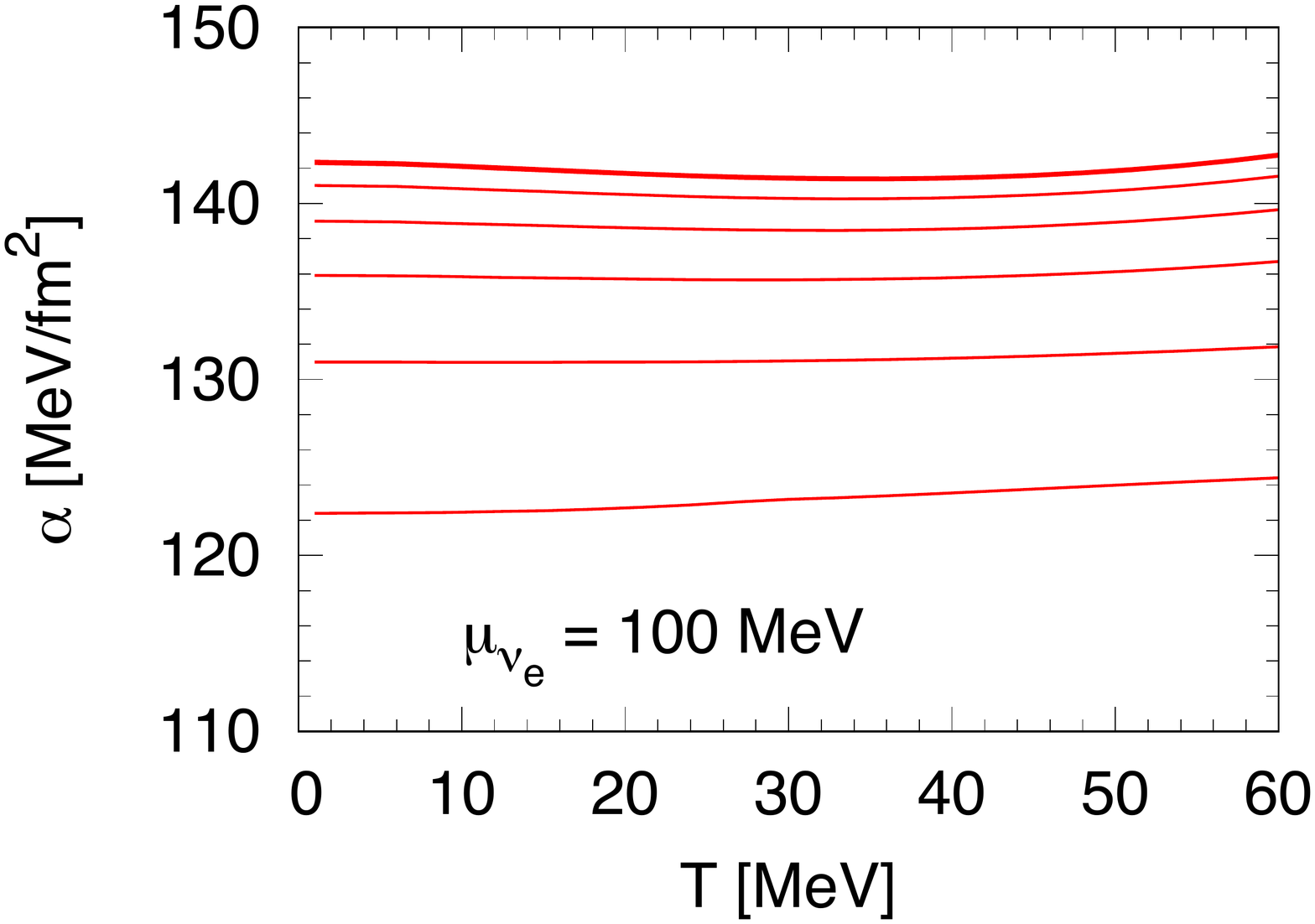}
\vskip -5mm
\caption{Total surface tension $\alpha$ as a function of temperature $T$ for different droplet radii $R$. 
From bottom to top we have $R \, [\textrm{fm}]= 2, 5, 10, 20, 50, 500, \infty$.  
From numerical simulations of protoneutron star evolution, it is known that chemical potential of trapped neutrinos $\mu_{{\nu}_e}$ decreases from $\sim 200$ MeV to essentially zero in $\sim 1$ minute. Here we adopt two representative values, $\mu_{{\nu}_e}=0$ and 100 MeV. The baryon number density of each point of the curves cannot be chosen arbitrarily but must be  determined consistently from Gibbs conditions and flavor conservation with respect to a hadronic equation of state (in this case obtained from a relativistic non-linear Walecka model) \cite{Lugones2011}. 
Note that $\alpha$ is quite independent of $T$ but there is a significant dependence with $R$. For quark matter we use the NJL model. For more details see Ref. \cite{Lugones2011}.  }
\label{fig:11}
\end{figure*}

\begin{figure*}[tbh]
\begin{center}
\includegraphics[angle=0,scale=0.19]{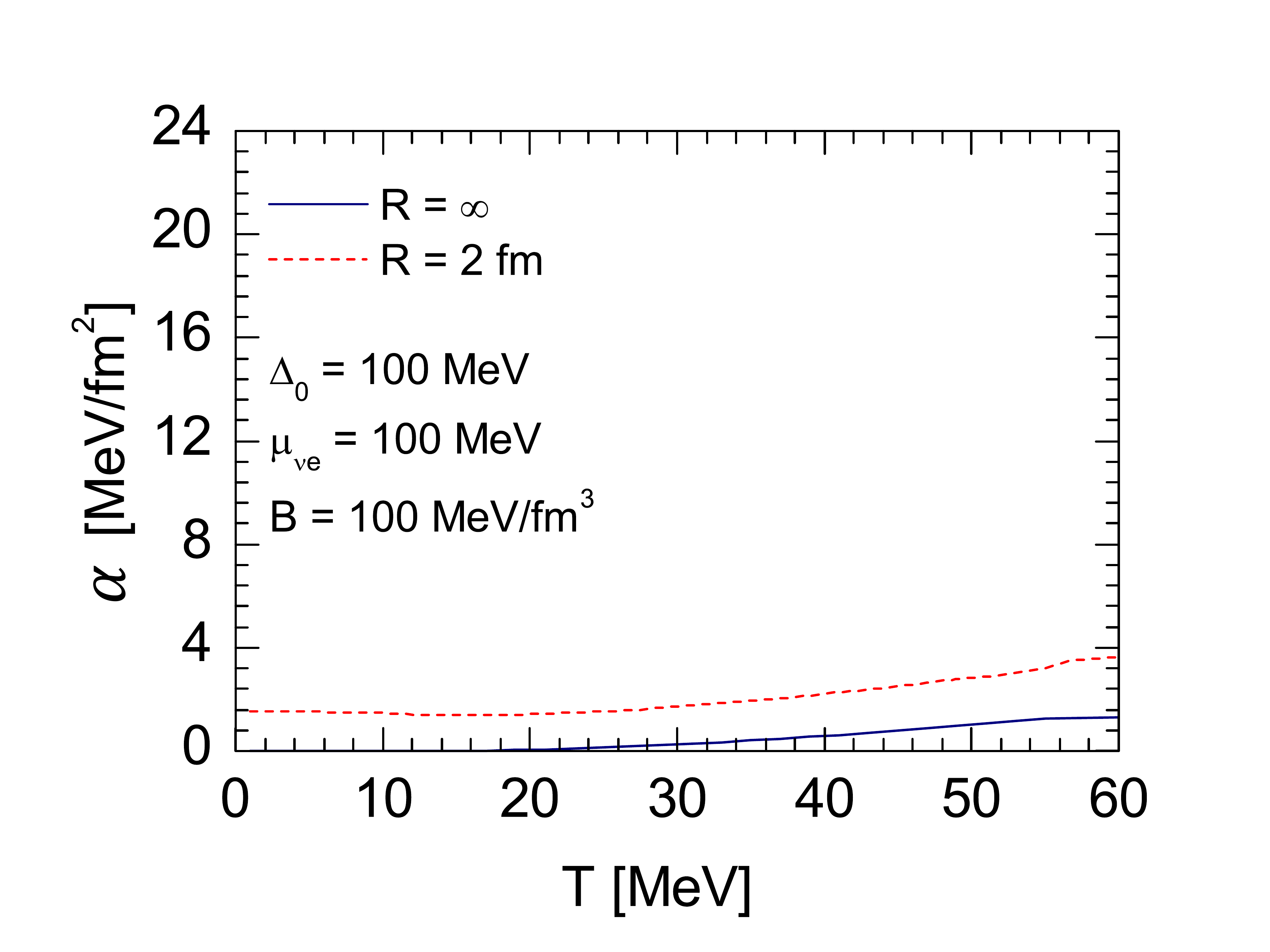}
\includegraphics[angle=0,scale=0.19]{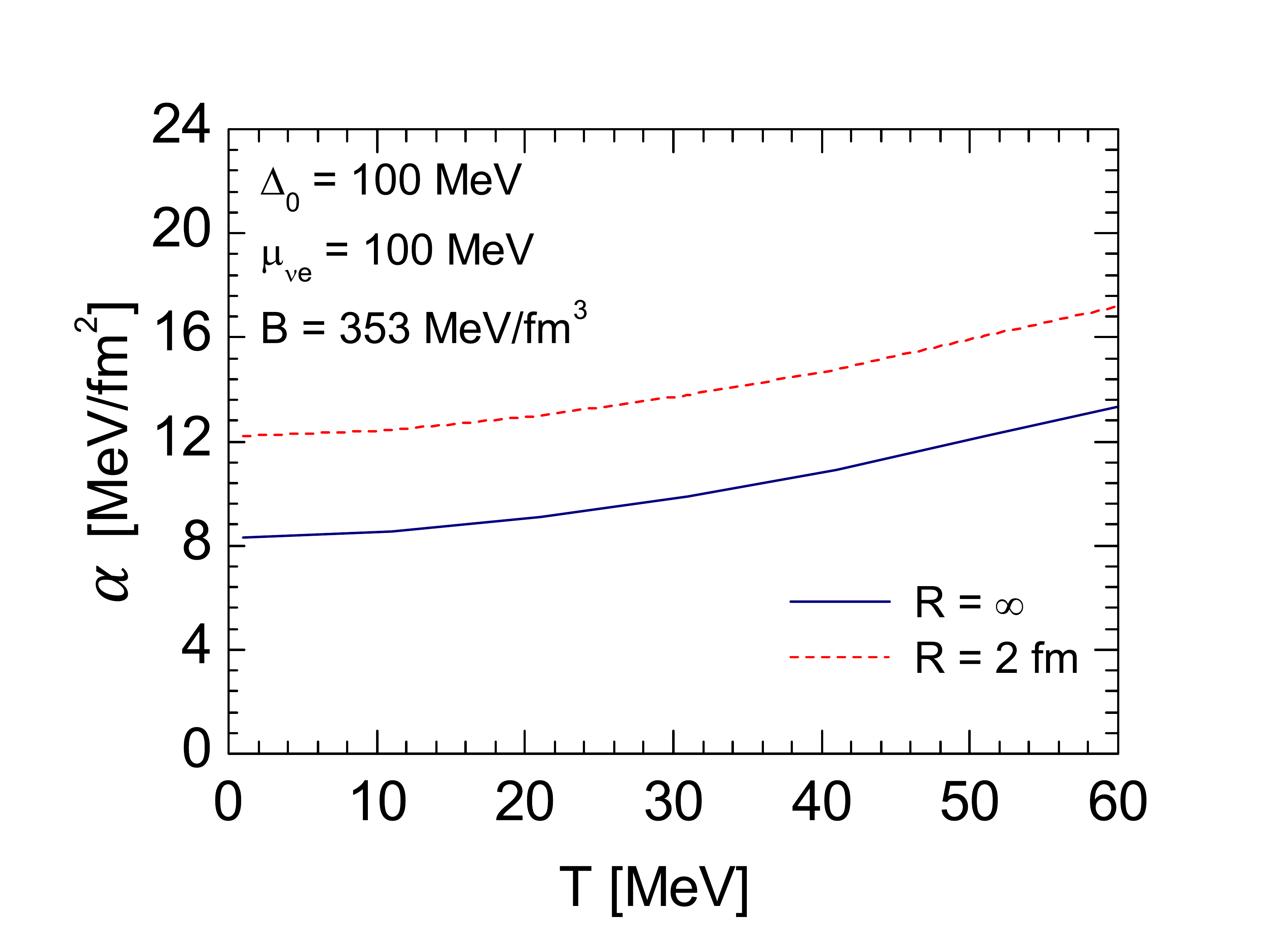}
\end{center}
\vskip -5mm
\caption{Same as in the previous figure but using the MIT bag model and  $R= 2, \infty$ fm. The pairing gap has a temperature dependence $\Delta(T) = \Delta_0 (1 - T^2 / T_c^2)^{1/2}$ with $\Delta_0= 100$ MeV and $T_c= 0.57 \Delta_0$. We use two values of the bag constant $B$. For more details see Ref. \cite{doCarmo2013}.}
\label{fig:12}
\end{figure*}

In proto NSs, matter can reach temperatures of the order of $50$ MeV and the neutrino chemical potential is very high because neutrinos are trapped due to their short mean free path. In such conditions, one may wonder whether deconfined quark matter drops can nucleate within an initially hadronic NS. To study this, we have employed a two phase description of the first order hadron-quark phase transition.
For the hadronic phase we used the nonlinear Walecka model, which includes the whole baryon octet, electrons, and trapped electron neutrinos in equilibrium under weak interactions. For the just deconfined quark matter we used an $SU(3)_f$ NJL model including color superconducting
quark-quark interactions and the MIT bag model \cite{Lugones2011,doCarmo2013}. In the quark phase, finite size effects are included via the MRE formalism. 
We assume that quark droplets can be formed if Gibbs conditions are fulfilled (i.e. if the Gibbs free energy of both phases is the same at the same pressure and temperature). Since deconfinement is driven by strong interactions, it is reasonable to assume that the just formed quark phase has initially the same flavor composition as the hadronic-stable phase from which it has been produced. Thus, these drops are initially  out of chemical equilibrium under weak interactions. In such thermodynamic conditions, we have calculated the surface tension $\alpha$ of quark matter. Our results are displayed in  Figs. \ref{fig:11} and \ref{fig:12}. We see that the NJL model gives larger values for $\alpha$ than the MIT bag model; typically more than one order of magnitude.

\section{Surface tension and the quark-hadron interface in hybrid NS\lowercase{s} }

In hybrid stars, the quark-hadron interface can be either a sharp discontinuity or a mixed phase where the electric charge is zero globally but not locally, and therefore charged hadronic and quark matter may share a common lepton background, leading to a quark-hadron mixture extending over a wide density region of the star. The actual outcome depends crucially on the surface tension $\alpha$ of three-flavor quark matter in equilibrium under weak interactions.  To address this issue, we include the effect of color superconductivity (in the 2SC phase) within the NJL model and describe finite size effects within the MRE framework \cite{Mudhahir 2013}.

\begin{figure}[tb]
\centering
\vskip  -5mm
\includegraphics[angle=0,scale=0.30]{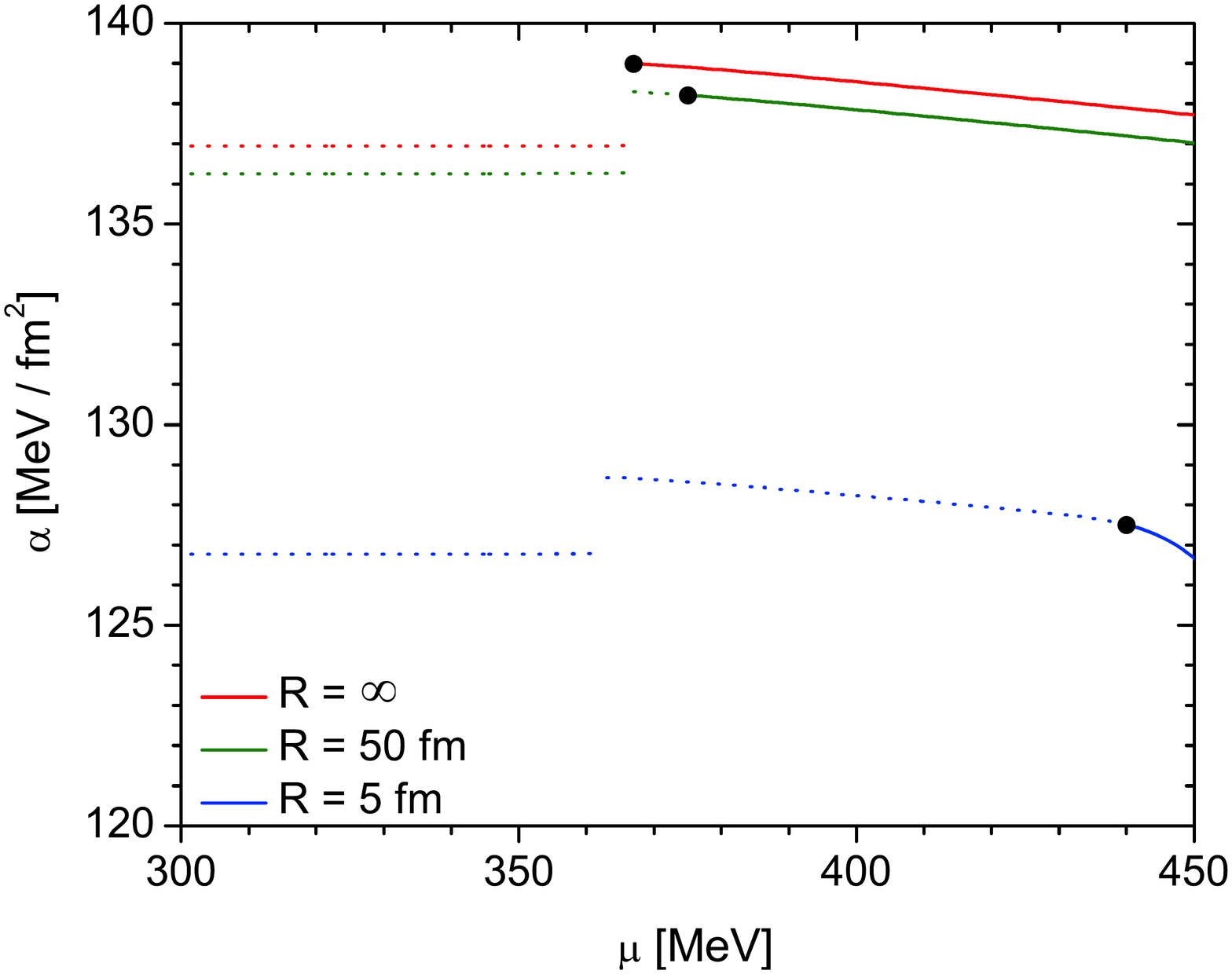}
\caption{Surface tension for cold quark matter in chemical equilibrium under weak interactions using the NJL model. We obtain large values of $\alpha$ suggesting that a mixed phase would be precluded at hybrid stars. For more details see \cite{Mudhahir 2013}.}
\label{fig:2}
\end{figure}

In Fig. \ref{fig:2} we show $\alpha$  as a function of the quark chemical potential for  $T=0$ MeV, which is representative of the conditions prevailing in old NS. We show results for drops with radii ranging from very small values of 5 fm, which have a large energy cost due to surface and curvature effects, to the bulk limit of $R = \infty$. The left branches correspond to the chiral symmetry broken phase and the right curves after the discontinuity to the 2SC phase (please see \cite{Mudhahir 2013} for the discussion about the curves at the left of the dot).

The large values of $\alpha$ have strong consequences for the physics of neutron star interiors, because the energy cost of forming quark drops within the mixed phase of hybrid stars would be very large.  According to \cite{Voskresensky2003Tatsumi2003Maruyama2007Endo2011}, beyond a limiting value of $\alpha \approx 65$ MeV/fm$^2$ the structure of the mixed phase becomes mechanically unstable  and local charge neutrality is recovered. Therefore, our results indicate that the hadron-quark inter-phase within a hybrid star should be a sharp discontinuity.

\section{Highly magnetized degenerate quark matter}

\begin{figure}[tb]
\includegraphics[angle=0,scale=0.4]{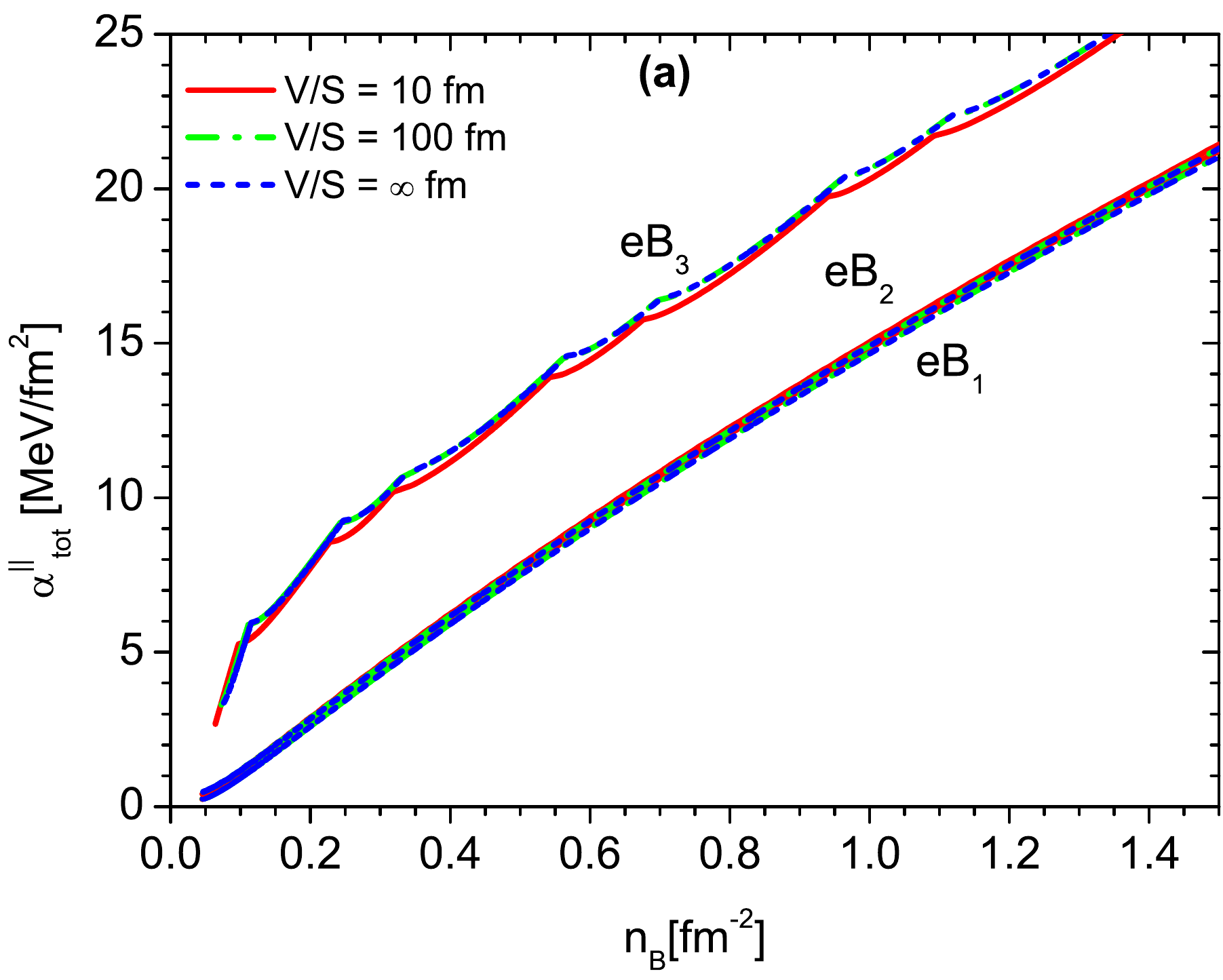}
\includegraphics[angle=0,scale=0.4]{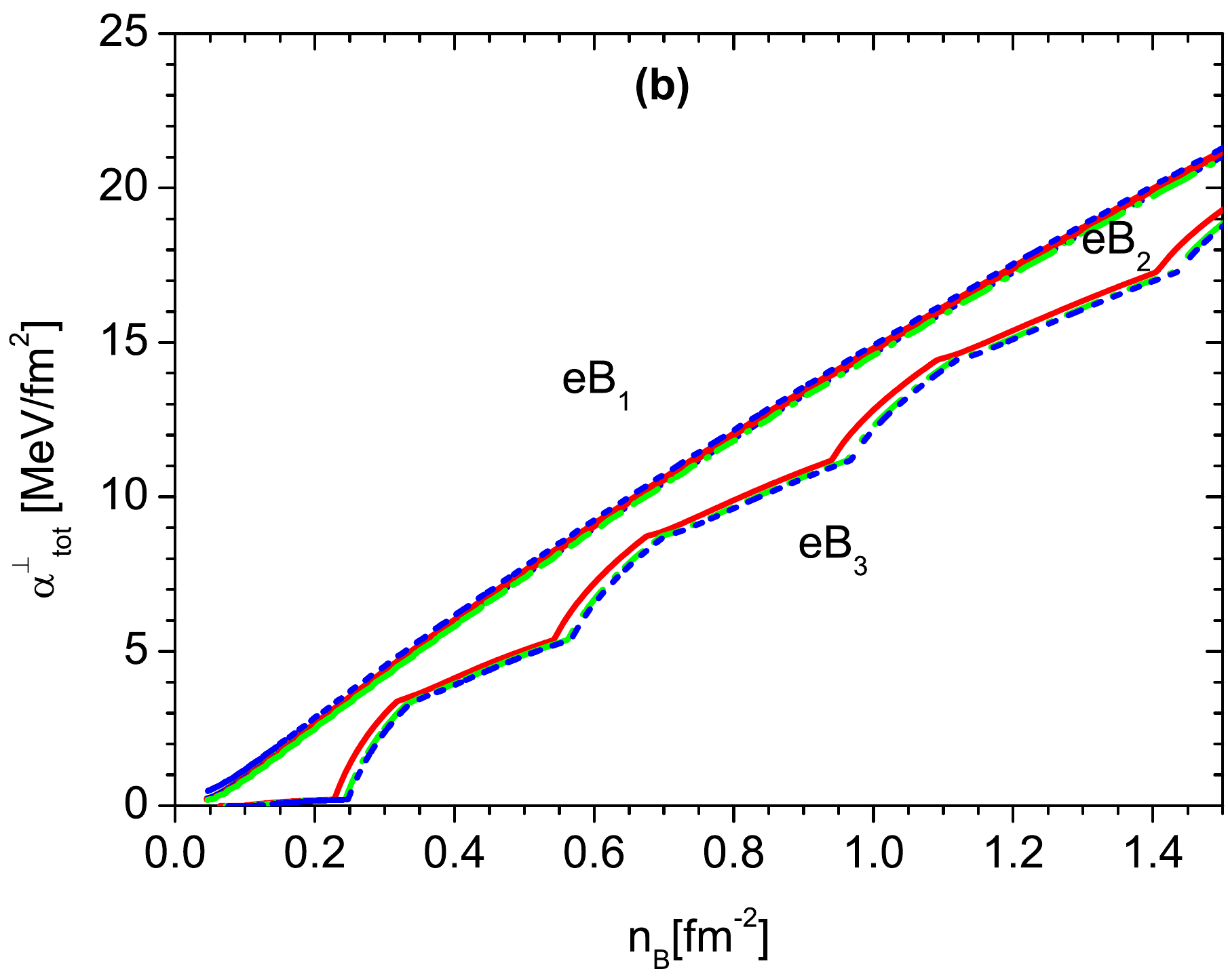}
\caption{Total surface tension at zero temperature as a function of the baryon number density, for different values of $V/S$. For the magnetic field we use $e B_1 = 5 \times 10^{-4}$ GeV$^2$, $e B_2 = 5 \times 10^{-3}$ GeV$^2$ and $e B_3 = 5 \times 10^{-2}$ GeV$^2$ (1 GeV$^2 = 1.690 \times 10^{20}$ G). For more details see  \cite{German2017}.}
\label{fig:4}
\end{figure}

Systems of strongly interacting matter under the influence of intense magnetic fields are the subject of current studies. They have direct application to the physics of NSs and the properties of the quark gluon plasma produced in relativistic heavy-ion collisions. Here we study the surface tension of magnetized quark matter within the MRE. Quark matter is described as a mixture of free Fermi gases composed by quarks $u, d, s$ and electrons in chemical equilibrium under weak interactions \cite{German2017}. Due to the presence of strong magnetic fields, the transverse motion of these particles is quantized into Landau levels and, as a consequence, the surface tension has a different value in the parallel and transverse directions with respect to the magnetic field.
We calculate the surface tension in the longitudinal and transverse directions for quark-matter drops with different sizes $V/S$ (volume/surface). Due to the magnetic field effect, such drops are expected to be prolate. However, $\alpha_f^{\parallel}$ and $\alpha_f^{\perp}$ (parallel and perpendicular to the magnetic field) depend on the shape of the drop only through the ratio $V/S$. In the present work, $V/S$ is taken as a free parameter with values 10 fm, 100 fm, and $\infty$. Our results are shown in Fig. \ref{fig:4}.
The largest contribution to the surface tension comes from strange quarks which have a surface tension an order of magnitude larger than the surface tension for $u$ or $d$ quarks and more than two orders of magnitude larger than for electrons. The parallel surface tension increases with the magnetic field and the transverse one decreases.

These results show that, although the magnetic field can change $\alpha$ significantly, its effect would not be large enough to change the nature of the hadron-quark interface from mixed to sharp \cite{German2017}. However, since quark-matter drops adopt a prolate shape inside a strong magnetic field, we may expect significant changes in the geometrical structure of the drops, rods, slabs, and all the “pasta-phase” configurations expected to exist if global charge neutrality is allowed.

\section{Finite size effects at zero chemical potential from the Polyakov loop NJL model}

Finite size effects in strongly interacting matter are also relevant in the context of heavy ion collisions and in the early Universe. Although lattice QCD methods had a huge progress in recent years, it is extremely difficult to solve QCD in the regime of intermediate temperatures and chemical potentials. 
Additionally, even at $\mu=0$ where lattice QCD calculations are feasible, studies of finite size effects from first principles are not available.  
In this context, effective models are very useful for describing the behavior  of strongly interacting matter. In a recent work \cite{polyakov}, we studied finite size effects within the Polyakov loop Nambu-Jona-Lasinio model (PNJL) for two light and one heavy quarks at vanishing baryon chemical potential and finite temperatures. We included different Polyakov loop potentials and checked that the predictions of our effective model in bulk are in agreement with lattice QCD results. Finite size effects were  included through the MRE formalism.

\begin{figure}[tb]
\includegraphics[angle=0,scale=0.33]{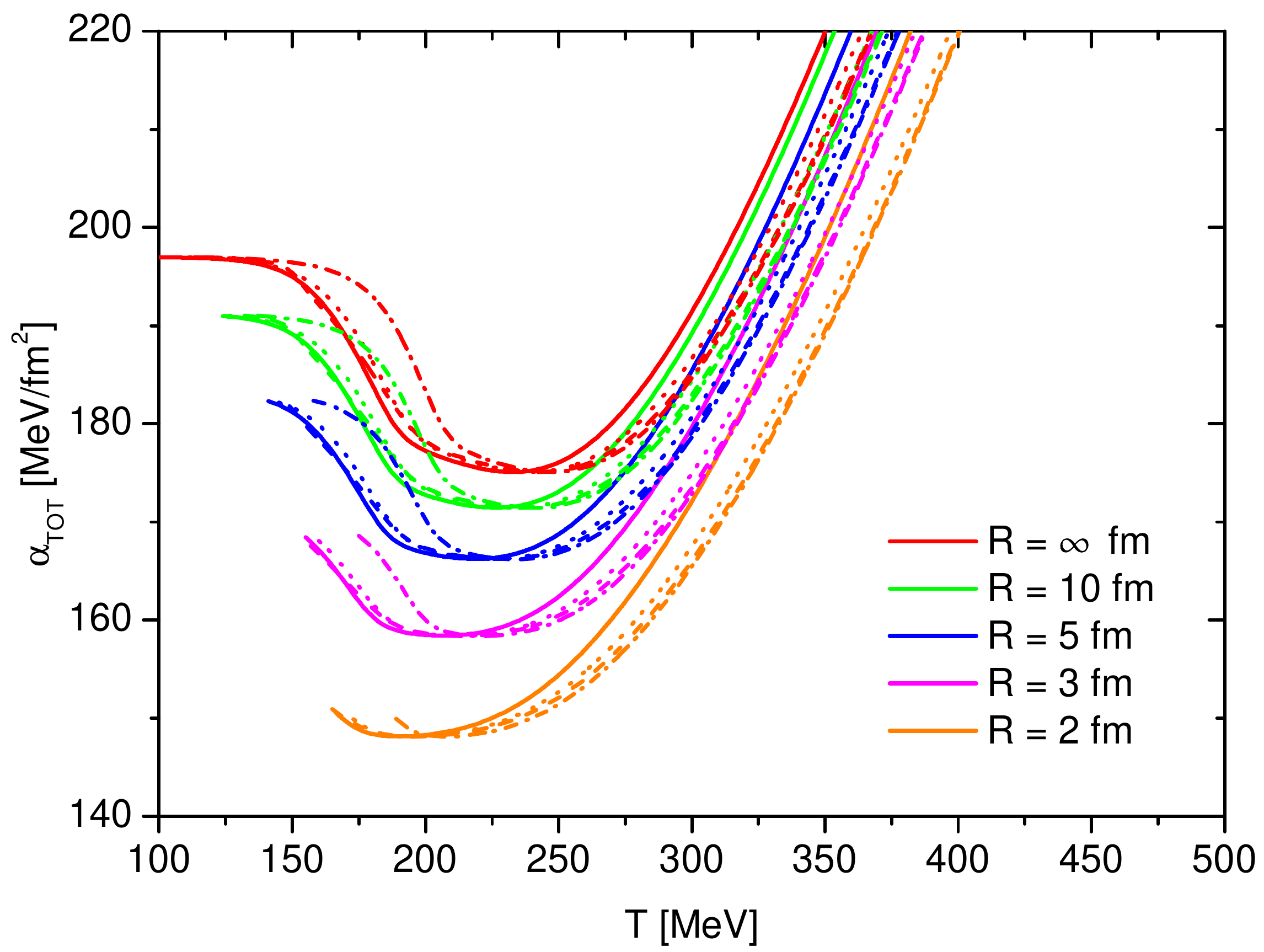}
\caption{Total surface tension as a function of temperature at zero chemical potential. We display results for different Polyakov loop potentials in solid, dashed, dotted and dashed-dot lines (see \cite{polyakov} for details).}
\label{fig:5}
\end{figure}

In Fig. \ref{fig:5} we show the total surface tension $\alpha_{\mathrm{TOT}}$ for drops with different sizes, where $\alpha_{\mathrm{TOT}}= \sum_i \alpha_i$  includes the contribution of $u$, $d$ and $s$ quarks.  
We have checked that $\alpha_{s}$ is more than $10$ times larger than $\alpha_{u}$ and $\alpha_{d}$, in qualitative accordance with results for cold quark matter at very high densities \cite{Mudhahir 2013,German2017} that show that the  total surface tension $\alpha_{\mathrm{TOT}}$ is largely dominated by the contribution of $s$ quarks. 
The surface tension shows a significant variation with $R$ at all temperatures, specially for small drops with radii below 10 fm. There is also a considerable dependence on the Polyakov loop potential. 
At large temperatures, $\alpha_{\mathrm{TOT}}$ increases monotonically with $T$. At lower $T$, $\alpha_{\mathrm{TOT}}$ has a local minimum associated with the chiral transition of strange quarks.  
At temperatures below the minimum, $\alpha_{\mathrm{TOT}}$ tends to a constant value which is of the same order of the values obtained within the NJL model for cold quark matter ($T=0$) at finite chemical potentials ($\mu = 0-450$ MeV). 
%
\section{Conclusions}

(1) Surface tension in drops nucleated out of chemical equilibrium at proto NS conditions is at least ten times larger within the NJL model than within the MIT bag model.

(2) Within the NJL model, surface tension in cold hybrid star conditions is  $\sim 150$ MeV/fm$^2$.  For such large values, mixed phases are mechanically unstable and the hadron-quark interface in a hybrid star should be a sharp discontinuity. 

(3) Within the MIT bag model, the surface tension of highly magnetized quark matter is in the range $2-20$ MeV/fm$^2$ for a wide range of baryon number densities. The largest contribution comes from $s$ quarks and the results are quite insensitive to the size of the drop. For $eB \gtrsim 5 \times 10^{-3}$ GeV$^2$, there is a significant increase in the  surface tension parallel to the magnetic field and a significant decrease in the one perpendicular to $B$, with respect to the unmagnetized case.
For these values of $\alpha$, a mixed phase is expected within a hybrid magnetar.

(4) Surface tension within the PNJL model at vanishing chemical potential is dominated by the contribution of $s$ quarks,  it changes significantly with $R$ and there is some dependence on the choice of the Polyakov loop potential. At large temperatures, $\alpha$ increases monotonically with  $T$; at very low $T$ it saturates to a constant value.



\begin{thebibliography}{99}

\bibitem{Buballa2013Pinto2012Wen2010Palhares2010}
M. Buballa and S. Carignano, Phys. Rev. D 87, 054004 (2013). M. B. Pinto, V. Koch, and J. Randrup, Phys. Rev. C 86, 025203
(2012).  G. Lugones,   Eur. Phys. J. A  52,  53 (2016).


\bibitem{Lugones2011} G. Lugones and A. G. Grunfeld, Phys. Rev. D 84, 085003 (2011).

\bibitem{doCarmo2013} T. A. S. do Carmo, G. Lugones, and A. G. Grunfeld, J. Phys. G: Nucl. Part. Phys. 40, 035201 (2013).

\bibitem{Voskresensky2003Tatsumi2003Maruyama2007Endo2011}
D. N. Voskresensky, M. Yasuhira, and T. Tatsumi, Nucl. Phys.
A 723, 291 (2003). T. Maruyama, S. Chiba, H.-J. Schulze, and T. Tatsumi, Phys.Rev. D 76, 123015 (2007). 

\bibitem{Jaikumar2006}
P. Jaikumar, S. Reddy, and A. W. Steiner, Phys. Rev. Lett. 96,
041101 (2006).

\bibitem{Alford2006}
M. G. Alford, K. Rajagopal, S. Reddy, and A. W. Steiner, Phys.
Rev. D 73, 114016 (2006).

\bibitem{Berger1987}
M. S. Berger and R. L. Jaffe, Phys. Rev. C 35, 213 (1987).

\bibitem{Heiselberg1993Iida1998}
H. Heiselberg, C. J. Pethick, and E. F. Staubo, Phys. Rev. Lett.
70, 1355 (1993). K. Iida and K. Sato, Phys. Rev. C 58, 2538 (1998).

\bibitem{Bombaci2007Bombaci2009}
I. Bombaci, G. Lugones, and I. Vidana, ˜ Astron. Astrophys. 462,
1017 (2007). I. Bombaci et al., Phys. Lett. B 680, 448 (2009).

\bibitem{Alford2001}
M. G. Alford, K. Rajagopal, S. Reddy, and F. Wilczek, Phys.
Rev. D 64, 074017 (2001).

\bibitem{Madsen-dropKiriyama1Kiriyama2}
J. Madsen, Phys. Rev. D 50, 3328 (1994). O. Kiriyama and A. Hosaka, Phys. Rev. D 67, 085010 (2003). 
 
\bibitem{Mudhahir 2013}
G. Lugones, A. G. Grunfeld, and M. A. Ajmi, Phys. Rev. C 88,
045803 (2013).


\bibitem{German2017} G. Lugones and A. G. Grunfeld, Phys. Rev. C 95, 015804 (2017).

\bibitem{polyakov} A. G. Grunfeld and G. Lugones, arXiv:1711.07559.

\end{thebibliography}
\end{document}